\documentclass{elsart}
\makeatletter
\def\fps@figure{p}
\def\fps@table{p}
\makeatother
\usepackage{graphicx}
\newcommand\UT{University of Tokyo}
\newcommand\hongo{7-3-1~Hongo, Bunkyo-ku, Tokyo~113-0033, Japan}
\newcommand\gagg{g_{a\gamma\gamma}}

\begin{document}
\begin{frontmatter}
%
%
\title{Search for sub-electronvolt solar axions
  using coherent conversion of axions into photons
  in magnetic field and gas helium}
%
%
\author[ICEPP,RESCEU]{Yoshizumi Inoue\corauthref{AUTH}},
\corauth[AUTH]{Corresponding author}
\ead{berota@icepp.s.u-tokyo.ac.jp}
\author[ICRR]{Toshio Namba},
\author[ICRR]{Shigetaka Moriyama},
\author[PHYS,RESCEU]{Makoto Minowa},
\author[PHYS]{Yuko Takasu},
\author[PHYS]{Takashi Horiuchi},
\and
\author[KEK]{Akira Yamamoto}
\address[ICEPP]{International Center for Elementary Particle Physics,
  \UT, \hongo}
\address[PHYS]{Department of Physics, School of Science, \UT, \hongo}
\address[ICRR]{Kamioka Observatory, Institute for Cosmic Ray Research, \UT,
  Kamioka-cho, Yoshiki-gun, Gifu~506-1205, Japan}
\address[KEK]{High Energy Accelerator Research Organization (KEK),
  1-1~Oho, Tsukuba, Ibaraki~305-0801, Japan}
\address[RESCEU]{Research Center for the Early Universe (RESCEU),
  School of Science, \UT, \hongo}
%
%
\begin{abstract}
A search for solar axions has been performed
using an axion helioscope
which is equipped with
a 2.3\,m-long 4\,T superconducting magnet,
PIN-photodiode X-ray detectors,
and a telescope mount mechanism to track the sun.
In addition, a gas container to hold dispersion-matching gas
has been developed and a mass region up to $m_a=0.27\rm\,eV$
was newly explored.
From the absence of any evidence,
analysis sets a limit on axion-photon
coupling constant to be
$\gagg<\mbox{6.8--10.9}\times10^{-10}\rm GeV^{-1}$
for the axion mass of
$0.05<m_a<0.27\rm\,eV$
at 95\% confidence level,
which is more stringent than the limit inferred from
the solar-age consideration.
This result
gives currently the most stringent observational limit
on the existence of solar axions
in this mass region.
\end{abstract}

%
%
\end{frontmatter}

%
%
\section{Introduction}

Quantum chromodynamics (QCD) is widely accepted as
the theory of strong interactions,
but has one flaw called the strong CP problem,
i.e., the neutron electric dipole moment is
$O(10^9)$ times smaller than expected.
The axion is a Nambu--Goldstone boson
of the broken Peccei--Quinn (PQ) symmetry
which was introduced to solve this problem
\cite{axion-bible1,axion-bible2,axion-bible3,axion-bible4,axion-bible5}.
The expected behavior of an axion is characterized
mostly by the scaling factor of the PQ symmetry breaking, $f_a$,
and so its mass, $m_a$, which is directly related to $f_a$ by
$m_a=6\times10^{15} [\mathrm{eV}^2]/f_a$.
Laboratory experiments, astrophysical and
cosmological considerations have constrained
\cite{axion-limit1,axion-limit2,axion-limit3}
the allowed mass region into two.
One is at micro- to millielectronvolts order,
where
axions can be an ideal dark matter,
and the other called `hadronic axion window'
is at around one to a few electronvolts,
where some hadronic axion models are possible.
In the latter region,
the sun can be a powerful source of axions
and the so-called `axion helioscope' technique
may enable us to detect such axions directly
\cite{sikivie1983,bibber1989}.

\begin{figure}
  \includegraphics[scale=0.8]{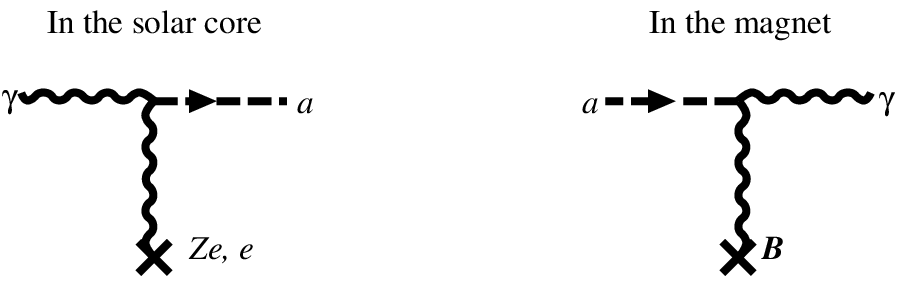}
  \caption{The solar axions produced via the Primakoff process
    in the solar core are, then, converted into X-rays
    via the reverse process in the magnet.}
  \label{fig:principle}
\end{figure}

The principle of the axion helioscope is illustrated
in Fig.~\ref{fig:principle}.
Axions would be produced
through the Primakoff process in the solar core.
The differential flux of solar axions
at the Earth is approximated by
\cite{bibber1989,solax1999}
\begin{eqnarray}
  \d \Phi_a/\d E&=&4.02\times10^{10}[\mathrm{cm^{-2}s^{-1}keV^{-1}}]
  \nonumber\\
  &&{}\times\left[\gagg\over10^{-10}\mathrm{GeV}^{-1}\right]^2
  {(E/1\,\mathrm{keV})^3\over
    \exp(E/1.103\,\mathrm{keV})-1},
  \label{eq:aflux}
\end{eqnarray}
where
$\gagg$ is the axion-photon coupling constant.
Their average energy is 4.2\,keV
reflecting the core temperature of the sun.
Then, they would be coherently converted into X-rays
through the inverse process
in a strong magnetic field at a laboratory.
The conversion rate is given by
\begin{equation}
  P_{a\to\gamma}
  ={\gagg^2\over4}\left|\int_0^L B_\bot\e^{\mathrm{i}qz}\d z\right|^2,
\end{equation}
where
$z$ is the coordinate along the incident solar axion,
$B_\bot$ is the strength of the transverse magnetic field,
$L$ is the length of the field along $z$-axis,
$q=|(m_\gamma^2-m_a^2)/2E|$ is the momentum transfer
by the virtual photon,
and
$m_\gamma$ is the effective mass of the photon
which equals zero in vacuum.

From 26th till 31st December 1997,
the first measurement \cite{sumico1997}
was performed
using an axion helioscope with a dedicated superconducting magnet
which we will describe in detail in Section \ref{sec:app}
except that the gas container was absent and
the conversion region was vacuum.
From the absence of an axion signal,
an upper limit on the axion-photon coupling is given to be
$\gagg<6.0\times10^{-10}\rm GeV^{-1}$ (95\% CL)
for $m_a<0.03\rm\,eV$.
However,
it was less sensitive to heavier axions,
since the momentum transfer, $q$, becomes non-negligible.

If one can adjust $m_\gamma$ to $m_a$,
coherence will be restored.
This is achieved by filling the conversion region with gas.
A photon in the X-ray region acquires a positive effective mass
in a medium.
In light gas,
such as hydrogen or helium,
it is well approximated by
\begin{equation}
  m_\gamma=\sqrt{4\pi\alpha N_e\over m_e},
\end{equation}
where $\alpha$ is the fine structure constant,
$m_e$ is the electron mass,
and $N_e$ is the number density of electrons.
We adopted cold helium gas as a dispersion-matching medium.
Here, light gas was preferred since it minimizes self absorption
by gas.
It is worth noting that
helium remains at gas state even at 5\,K,
the operating temperature of our magnet.
Since the bore of the magnet is limited in space,
the easiest way is to keep the gas
at the same temperature as the magnet.
Moreover,
axions as heavy as a few electronvolts
can be reached
with helium gas of only about one atmosphere
at this temperature.

In this paper, we will present the result of
a measurement
in which we scanned the mass region up to 0.27\,eV.

%
%
\section{Experimental apparatus}
\label{sec:app}

\begin{figure}
  \includegraphics[scale=0.8]{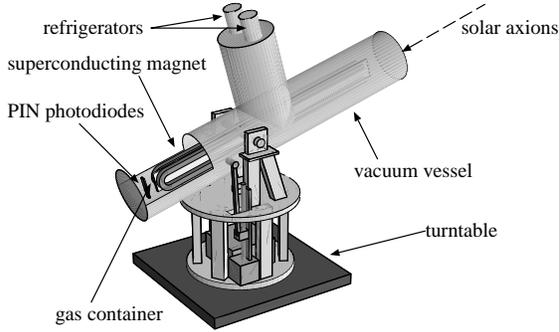}
  \caption{The schematic view of the axion helioscope.}
  \label{fig:sumico}
\end{figure}

The schematic figure of the axion helioscope is shown
in Fig.~\ref{fig:sumico}.
It is designed to track the sun in order to achieve
long exposure time.
It consists of
a superconducting magnet, X-ray detectors, a gas container,
and an altazimuth mounting.
In the following paragraphs,
we will describe each part in due order.

The superconducting magnet \cite{sato1997}
consists of two 2.3-m long
race-track shaped coils running parallel
with a 20-mm wide gap between them.
The magnetic field in the gap is 4\,T
perpendicular to the helioscope axis.
The coils are kept at 5--6\,K during operation.
In order to make it easy to swing this huge cryogenic apparatus,
two devices are engaged.
First, the magnet was made cryogen-free
by making two Gifford-McMahon refrigerators
to cool it directly by conduction.
Second, a persistent current switch was equipped.
Thanks to this, the magnet can be freed from
thick current leads after excitation,
and the magnetic field is very stable for a long period of time
without supplying current.

The container to hold dispersion-matching gas is inserted
in the $20\times92\,\mathrm{mm^2}$
aperture of the magnet.
Its body is made of four 2.3-m long 0.8-mm thick
stainless-steel square pipes
welded side by side to each other.
The entire body is wrapped with 5N high purity
aluminium sheet to achieve high uniformity of temperature.
The measured thermal conductance between the both ends was
$1\times10^{-2}\mathrm{W/K}$ at 6\,K.
One end at the forward side of the container is sealed
with welded plugs
and is suspended firmly by three Kevlar cords.
The opposite side nearer to the X-ray detectors
is flanged and fixed to the magnet.
At this end of the container, gas is separated from vacuum
with an X-ray window manufactured by METOREX
which is transparent to X-ray above 2\,keV
and can hold gas up to 0.3\,MPa at liquid helium temperature.

Sixteen PIN photodiodes, Hamamatsu Photonics S3590-06-SPL,
are used as the X-ray detectors,
whose chip sizes are $11\times11\times0.5\rm\,mm^3$ each.
The effective area of a photodiode was measured
formerly using a pencil-beam X-ray source,
and found to be larger than $9\times9\,\mathrm{mm^2}$.
Each chip is mounted on a Kapton film
bonded to an Invar plate with cryogenic compatible adhesive.
The X-ray detectors
are mounted in a 10-mm thick radiation shielding box made of
oxygen-free high conductivity copper (OFHC Cu),
which is then surrounded by a lead shield of about 150\,mm thick.
The copper shield is operated at about 60\,K,
so that
it also functions as a cold finger for the X-ray detectors.
Details on the X-ray detector are given in Ref.\ \cite{naniwaPIN}.

The output from each photodiode is fed
to a charge sensitive preamplifier whose first-stage
FET is at the cryogenic stage near the photodiode chip
and the preamplifier outputs are digitized using
CAMAC flash analog-to-digital convertors (FADC's), REPIC RPC-081's,
at a sampling rate of 10 MHz.
The preamplifier outputs are also fed to
shaping amplifiers, Clear Pulse CP4026,
whose outputs are then discriminated to generate triggers.
Thus, waveforms of the sixteen preamplifier outputs
are recorded simultaneously
over 50 $\mu s$ before and after each trigger
to be committed to later off-line analysis.
Each detector was calibrated by 5.9-keV Mn X-rays
from a \nuc{55}{Fe} source installed in front of them.
The source is manipulated from the outside
and is completely retracted behind the shield
during the axion observations.

The entire axion detector is constructed in a vacuum vessel
and the vessel is mounted on an altazimuth mount.
Its trackable altitude ranges from $-28^\circ$ to $+28^\circ$
and its azimuthal direction is limited only by a limiter
which prevents the helioscope from endless rotation.
This view corresponds to an exposure time
of about a half of a day in observing the sun in Tokyo
at 139$^\circ$ 45' 48'' E and 35$^\circ$ 42' 49'' N.
This is enough for our purpose,
since background is measured during the other half of a day.
This helioscope mount is driven by two AC servo motors
controlled by a computer (PC).
The PC also monitors the azimuthal and altitudinal directions
of the helioscope regularly
by two precision rotary encoders
and forms a feedback controlling loop as a whole.
The U.S. Naval Observatory Vector Astronomy Subroutines (NOVAS) \cite{novas}
were used to calculate the solar position.
The directional origin was determined
using a theodolite.
The altitudinal origin was determined from a spirit level.
While the sun is not directly visible from the laboratory
in the basement floor,
the azimuthal origin was first determined from the observed
direction of Polaris ($\alpha$-UMi) outdoors,
and then it was introduced to the laboratory with the theodolite.

Since the effective aperture of the helioscope is narrow,
it is crucial to determine its accurate geometry.
The axis of the helioscope is defined by
two cross hairs at the edge of the vacuum vessel.
The position of each part of the helioscope was measured
relative to these cross hairs
from their exterior
using the theodolite
when they were installed.
The positions of the PIN photodiodes were determined relative to
the copper shielding box from a photo image
taken prior to the installation.
As it is hard to estimate analytically
the effect of the geometrical errors
as well as the effect of the size of the axion source,
we performed a Monte Carlo simulation and
found that the overall effective area is larger than $602\,\mathrm{mm}^2$
at 99\% confidence level.

%
%
\section{Measurement and Analysis}


From 29th July till 1st September 2000,
a measurement employing dispersion-matching gas was
performed
for ten photon mass settings
to scan up to 0.27\,eV
which is shown in Table \ref{tab:settings}.
The actual gas density was determined
using a Yokokawa MU101 precision pressure gauge
and two Lakeshore CGR thermistors
attached to the gas container,
and was calculated
based on interpolation of the tables from NIST
\cite{nist}.

Before obtaining energy spectra,
each event was categorized into two major groups,
the solar observation and the background.
Events while the measured direction agreed with the sun
are counted as former.
When the sun is completely out of the magnet aperture,
events are counted as latter.
Otherwise events are discarded.

Then, the following bad events are discarded;
(a) events containing saturated waveforms,
(b) events containing multiple pulses,
(c) events containing pulses which are too early or too late
compared to their triggers,
and
(d) events containing pulses whose rise-time is too slow.
Cut (a) can be classified into two cases.
One is such events where the saturation is caused by a giant pulse,
e.g., by cosmic muons.
Another is a group of events in which microphonic noises were
so large that the baseline went out of the input range
of the FADCs.
They are distinguished by their waveforms
and
the live time was estimated after the most conservative manner.
No correction is needed for cut (b) which is also aimed at cosmic-ray events,
since the expected rate of
the axion events is extremely low.
Cut (c) and (d) were introduced to simplify the later
waveform analysis.
The loss in efficiency by cut (c) was found to be negligible.
The causes of the slow pulses may be some kind of
spontaneous noises of the detector,
but are yet to be identified.
However, it was confirmed that they were at least not X-rays,
since we could hardly find such events
in the data in which the \nuc{55}{Fe} source
was observed.
In order to estimate the loss in efficiency by cut (d),
the same cut was applied to
the \nuc{55}{Fe} source data.
We found it to be less than 2.6\%
using the worst value of the sixteen channels.

\begin{figure}
  \includegraphics{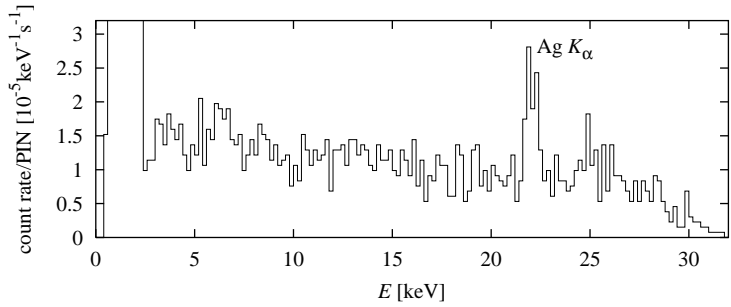}
  \caption{The background spectrum of the X-ray detector.
    The sixteen PIN photodiodes and the ten gas densities
    are combined.}
  \label{fig:allbg}
\end{figure}

We performed numerical pulse shaping to the raw waveforms
using the Wiener filter.
The energy of an X-ray is given by the peak height of
a wave after shaping.
The shaped waveform is given by
\begin{equation}
  \label{eq:wiener}
  U(\omega)=
  {S^\ast(\omega) C(\omega)
    \over|N(\omega)|^2},
\end{equation}
where $U(\omega)$, $S(\omega)$, $C(\omega)$,
and $N(\omega)$ are Fourier transformations of
the shaped waveform,
the ideal signal waveform,
the measured waveform,
and the noise, respectively.
Noises are obtained by gathering waveforms
while no trigger exists,
and the ideal signal waveform is
approximated by averaging signals from 5.9-keV X-rays.
The response function of this waveform analysis,
i.e., non-linearity, gain walk by trigger timing, etc.,
was investigated thoroughly using simulated pulses
which were obtained by adding
the template waveform to the noise waveforms.
A correction was made
based on this numerical simulation.
Saturation arised at about 25\,keV,
therefore, $E>20\,\mathrm{keV}$ was not used
in the later analysis.
Fig.\ \ref{fig:allbg} shows the comprehensive background
spectrum of all the photodiodes and all the gas settings.
The background level was
about $1.5\times10^{-5}\mathrm{s^{-1}keV^{-1}/PIN}$
at $E=5\hbox{--}10\,\mathrm{keV}$.
The peak appearing at 22\,keV
is consistent with the
fluorescent X-rays from the small amount
of silver epoxy between the photodiode chip and the Kapton base.
By analysing the calibration data,
we found the energy resolution of each photodiode
to be
0.70--0.87\,keV (FWHM)
for 5.9-keV photons.

\begin{figure}
  \hbox{%
    \includegraphics{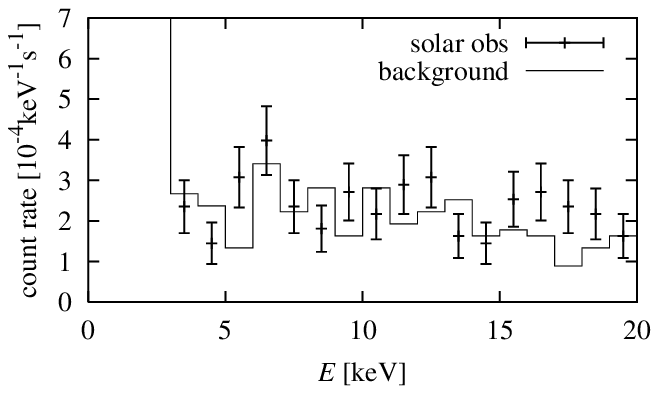}%
    \hskip 1cm
    \includegraphics{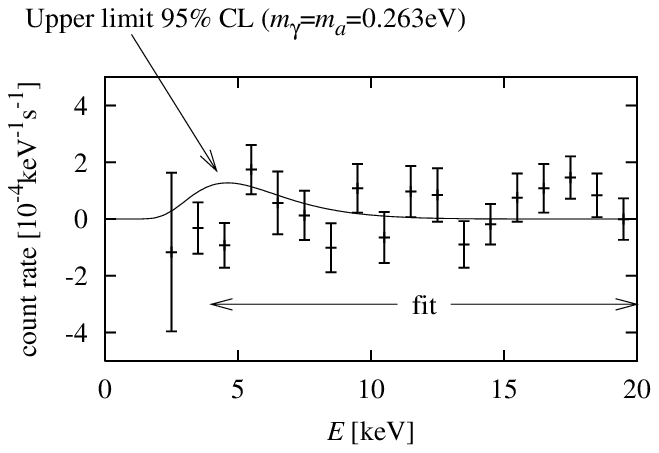}}
  \caption{The left figure shows the energy spectrum
    of the solar observation (error bars) and
    the background spectrum (solid line) when
    the gas density was tuned to $m_\gamma=0.263\rm\,eV$.
    The right figure shows the net energy spectrum of the left
    where the background is subtracted from the solar observation.}
  \label{fig:spec}
\end{figure}

In Fig.~\ref{fig:spec},
one of the energy spectra of the solar observation
is shown together with the background spectrum.
We searched for expected axion signals
which scale with $\gagg^4$
for various $m_a$
in these spectra.
The smooth curve in the figure represents an example for
the expected axion signal
where $m_a=m_\gamma=0.263\rm\,eV$ and $\gagg=7.4\times10^{-10}\rm GeV^{-1}$,
which corresponds to the upper limit at $m_a=0.263\rm\,eV$
estimated as follows.


A series of least $\chi^2$ fittings was performed assuming various
$m_a$ values.
Data from the ten different gas densities
were combined by using the summed $\chi^2$ of the ten.
The energy region of 4--20\,keV was used for fitting
where the efficiency of the trigger system is almost 100\%
and the FADCs do not saturate.
The PIN photodiodes were assumed to
have inactive surface layers of 
$6.1\,\mu\mathrm{m}$ \cite{sumico1997}.
As a result, no significant excess was seen for any $m_a$,
and thus an upper limit on $\gagg$ at 95\% confidence level
was given following the Bayesian scheme.
Fig.~\ref{fig:exclusion} shows
the limit plotted as a function of $m_a$.
The previous limit and some other bounds are also
plotted in the same figure.
The SOLAX~\cite{solax1999} is a solar axion experiment
which exploits the coherent conversion
on the crystalline planes in a germanium detector.
The COSME~\cite{cosme2002} is the same kind of experiment
but with a different Ge detector at a different site.
Though their limits $\gagg<2.7\times10^{-9}\rm GeV^{-1}$
and $\gagg<2.78\times10^{-9}\rm GeV^{-1}$,
respectively,
are loose,
they are virtually independent of $m_a$
like the following two theoretical bounds.
It is also worth noting that they are observational.
The limit $\gagg<2.3\times10^{-9}\rm GeV^{-1}$ is
the solar limit inferred from the solar age consideration
and the limit $\gagg<1\times10^{-9}\rm GeV^{-1}$
is a more stringent limit reported
by Schlattl \etal ~\cite{schlattl1999}
based on comparison between
the helioseismological sound-speed profile
and
the standard solar evolution models with energy losses by solar axions.

%
%
\section{Discussion and Conclusion}
\begin{figure}
  \includegraphics{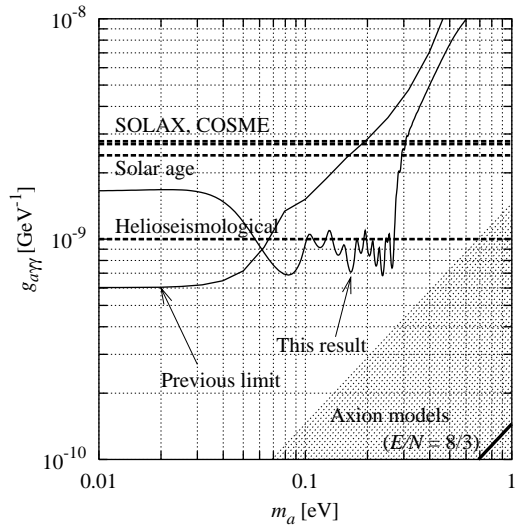}
  \caption{The exclusion plot on $\gagg$ to $m_a$
    is plotted
    where some other bounds are plotted together.
    The new limit and the previous one are plotted in solid lines.
    Dashed lines are
    the limit by SOLAX experiment,
    the limit by COSME experiment,
    the limit inferred from the solar age consideration,
    and the helioseismological bound.
    The four observational limits are at 95\% confidence level.
    The hatched area is the preferred axion models.
    The thick line corresponds to the case when
    a simple GUT is assumed.}
  \label{fig:exclusion}
\end{figure}

We have developed an axion helioscope
and
introduced cold helium gas
as the dispersion-matching medium
in the $4{\rm\,T}\times2.3\rm\,m$ magnetic field of the helioscope.
The axion mass up to 0.27\,eV has been scanned.
But no evidence for solar axions was seen.
A new limit on $\gagg$ shown in Fig. \ref{fig:exclusion}
was set for $0.05<m_a<0.27\rm\,eV$,
which is far more stringent than the solar-age limit.
This experiment could also clear some area beyond
the tighter helioseismological bound.
This limit assumes the standard solar model
by Eq.\ \ref{eq:aflux},
and depends especially on the core temperature.
Though the limit did not reach the preferred axion models,
this experiment is currently the only experiment
which has demonstrated enough sensitivity to detect such
solar axions that do not violate these solar limits.

The exact position of the `hadronic axion window'
or
whether it is still open or not are
controversial \cite{nomad2000,hisano1997}.
There is an argument that it has closed
by the globular-cluster limit \cite{gclimit}.
As for the solar limits,
Watanabe and Shibahashi \cite{watanabe2001} has recently argued
that the helioseismological bound
can be lowered to $\gagg<4.0\times10^{-10}\rm GeV^{-1}$
if the `seismic solar model' and
the observed solar neutrino flux are combined.
Discussions based on mutually independent assumptions
are meaningful.
And, it must be essential to test
by well-controlled experiments or by direct observations.
We are planning to upgrade our helioscope to scan
higher masses.
The CAST experiment \cite{satan1999} at CERN aims sensitivity as high as
$\gagg\approx5\times10^{-11}\mathrm{GeV}^{-1}$.

\begin{ack}
The authors thank the director general of KEK, Professor H. Sugawara,
for his support in the beginning of the helioscope experiment.
This research is supported by the Grant-in-Aid for COE research
by the Japanese Ministry of Education, Science, Sports and Culture,
and also by the Matsuo Foundation.
\end{ack}

%
%

%
%

\clearpage
%
%
\begin{table}
  \begin{center}
    \begin{tabular}{cccc}
      \hline
      molar density&
      $m_\gamma$&
      \multicolumn{2}{c}{live time [s]}\\
      $[\rm mol/m^3]$&
      $[\mathrm{eV}]$&
      \multicolumn{1}{c}{solar run}&
      \multicolumn{1}{c}{background}\\
      \hline
      $\phantom{0}4.137$&0.083&26245&35955\\
      $\phantom{0}8.325$&0.118&25866&34606\\
      $12.52\phantom{0}$&0.144&26111&35383\\
      $16.62\phantom{0}$&0.166&23088&34399\\
      $20.80\phantom{0}$&0.186&24876&35161\\
      $25.15\phantom{0}$&0.204&38355&50701\\
      $29.19\phantom{0}$&0.220&25478&35411\\
      $33.70\phantom{0}$&0.237&40356&51989\\
      $37.58\phantom{0}$&0.250&12638&30216\\
      $41.65\phantom{0}$&0.263&55289&67443\\
      \hline
    \end{tabular}
    \caption{Table of the gas settings and each live time.}
    \label{tab:settings}
  \end{center}
\end{table}

\begin{thebibliography}{99}
\bibitem{axion-bible1} R.D.~Peccei, H.R.~Quinn,
  Phys.\ Rev.\ Lett.\ 38 (1977) 1440.
\bibitem{axion-bible2} R.D.~Peccei, H.R.~Quinn,
  Phys.\ Rev.\ D 16 (1977) 1791.
\bibitem{axion-bible3} S.~Weinberg, Phys.\ Rev.\ Lett.\ 40 (1978) 223.
\bibitem{axion-bible4} F.~Wilczek, Phys.\ Rev.\ Lett.\ 40 (1978) 279.
\bibitem{axion-bible5} J.E.~Kim, Phys.\ Rep.\ 150 (1987) 1.

\bibitem{axion-limit1} M.S.~Turner, Phys.\ Rep.\ 198 (1990) 67.
\bibitem{axion-limit2} L.J~Rosenberg, K.~van~Bibber, Phys.\ Rep.\ 325 (2000) 1.
\bibitem{axion-limit3} G.G.~Raffelt, Phys.\ Rep.\ 333--334 (2000) 593.
  
\bibitem{sikivie1983} P. Sikivie, Phys.\ Rev.\ Lett.\ 51 (1983) 1415.

\bibitem{bibber1989} K. van Bibber \etal, Phys.\ Rev.\ D 39 (1989) 2089.

\bibitem{sumico1997} S. Moriyama \etal,
  Phys.\ Lett.\ B 434 (1998) 147.

\bibitem{naniwaPIN} T. Namba \etal,
  astro-ph/0109041, to be published in Nucl. Instr. Meth. A.

\bibitem{novas} G. H. Kaplan \etal,
  Astronomical Journal 97 (1989) 1197;\\
  URL: {\tt http://aa.usno.navy.mil/AA/software/novas/}.

\bibitem{solax1999} A.O. Gattone \etal,
  Nucl.\ Phys.\ B (Proc.\ Suppl.) 70 (1999) 59.

\bibitem{cosme2002} A. Morales \etal,
  Astropart.\ Phys.\ 16 (2002) 325.

\bibitem{schlattl1999} H. Schlattl, A. Weiss, G. Raffelt,
  Astropart.\ Phys.\ 10 (1999) 353.

\bibitem{watanabe2001} S. Watanabe, H. Shibahashi,
  hep-ph/0112012.

\bibitem{gclimit} G. G. Raffelt,
  Stars as Laboratories for Fundamental Physics,
  (The University of Chicago Press, 1996).

\bibitem{nomad2000} P.~Astier \etal,
  Phys.\ Lett.\ B 479 (2000) 371.

\bibitem{hisano1997} J. Hisano, K. Tobe, T. Yanagida,
  Phys.\ Rev.\ D (1997) 411.

\bibitem{satan1999} K. Zioutas \etal,
  Nucl. Instr. Meth. A 425 (1999) 480.

\bibitem{sato1997} Y. Sato \etal,
  Development of a Cryogen-free Superconducting Dipole Magnet,
  in: Proc. of the 15th International Conference on Magnet Technology (MT-15)
  (Beijing, October 1997),
  eds.: L. Liangzhen, S. Guoliao, Y. Luguang (Science Press, Beijing, 1998)
  pp. 262--265;
  KEK-Preprint-97-202 (November, 1997).

\bibitem{nist} Vincent D.~Arp and Robert D.~McCarty,
  Thermophysical Properties of Helium-4 from 0.8 to 1500 K
    with Pressures to 2000 MPa,
  NIST Technical Note 1334,
  (U.S. Department of Commerce,
  National Technical Information Service, 1989).
\end{thebibliography}
\end{document}